\documentclass[aps,prc,reprint,groupedaddress,showkeys,showpacs]{revtex4-1}
\usepackage{graphicx}
\usepackage{dcolumn}
\usepackage{bm}%
\usepackage{amsmath}
\usepackage{capt-of}
\usepackage{lineno,hyperref}

\begin{document}

\title{Effect of rotation and magnetic field in the gyroscopic precession around a neutron star}

\date{\today}

\author{Kamal krishna Nath}
\affiliation{Indian Institute of Science Education and Research Bhopal, Bhopal, India}
\author{Ritam Mallick}
\email{mallick@iiserb.ac.in}
\affiliation{Indian Institute of Science Education and Research Bhopal, Bhopal, India}

\begin{abstract}
We study the overall spin precession frequency of a test gyroscope around
a neutron star. The precession of the test gyroscope gives the signatures of the general relativistic effects that are present in the region of strong gravity of an NS. Using a numerical code, we find the precession of the test gyroscope for a rotating and a strongly magnetized NS. The magnetic
field distribution inside the NS is assumed either to be poloidal or toroidal. The overall spin precession rate is obtained by setting the orbital frequency of the gyroscope to a non-zero value but restricted to a timelike observer. The gyro frequency differs depending on the central object being a black hole or a neutron star. For neutron star, the gyro precession can even be calculated inside the star. We find that the gyroscope precession frequency depends on the star’s mass, rotation rate, and magnetic field configuration. 
 
\end{abstract}

\pacs{}

\keywords{}

\maketitle

\section{Introduction}

The present decade is the era of observational astronomy. The detection of gravitational waves has given a considerable impetus in the field of astronomy. It has revived the area of black holes (BHs) and neutron stars, the two most dense object in the universe. The observation of gravitational wave (GW) from BHs mergers (GW150914 \citep{abbott}) and subsequently the binary neutron star (NS) merger (GW170817 \citep{abbott1}) opened the field of multimessenger astronomy. These observations have given us probes by which we can further detect the property of these objects, which was previously clouded. The detections of NS merger has started to impose a severe constraint on the equation of state (EoS) of NS which was entirely arbitrary before \citep{annala,margalit,radice,ruiz,shibata,most}. With more astronomical observational satellites coming up (like LISA and Einstein telescope (for GW),Imaging X-ray Polarimetry Explorer (IXPE), Euclid (Dark Matter/Dark Energy (visible to near-infrared)), ATHENA (Advanced Telescope for High Energy Astrophysics)) our ability to understand the physics of such astrophysical objects will increase further.

The NS observation has seen significant and telling milestones recently,  starting from detection of two massive pulsars (PSR J1614-2230 and J2215+5135) \cite{demorest,antonoidis}) and the GW from NS merger\cite{abbott,abbott1}. Therefore, to understand the physics inside and near the NS, we should analyze them carefully. To achieve this goal, one must remember that NSs are dense objects and need General relativistic (GR) treatment. The spacetime inside and near an NS are curved and can be thought to be a theatre for such an event. The GR effect of these objects is revealed if we place a spinning gyroscope as a test particle near an NS or inside it. This gyroscope would feel both the geodetic effect (space-time curvature due to matter ) and also frame dragging due to the rotation of NS.

The study of gyroscope as a test particle is critical near an NS due to numerous reasons. The gyroscope will start to precess about its axis.
The test gyroscope in an actual scenario can be thought to be real particles inside the NS or near the NS in the accretion disk. The effect the gyroscope feels would be similar to the force that a particle would feel near an NS or even inside it. 

The study of gyro would help us in understanding the various aspect of NS: \\ 
a) the frame-dragging effect would induce an oblateness of the star and would also flatten the accretion disk.\\ 
b) The geodetic and frame-dragging effect would thus affect the constituent particle of the NS.\\ 
c) The theory of Quasi-periodic oscillations (QPOs) is not well understood, and such GR effect (as that felt by a gyroscope) can help us in understanding them in X-ray binaries.\\
d) GR effect can also have some impact on the moment of inertia of an NS; This would also ultimately have some effect on the generation of GW from NS.

In some NS the magnetic field is very strong, and they are now separately termed as magnetars \cite{duncan,thompson}. Some anomalous X-ray pulsars \cite{mereghetti,baykal1,baykal2} and some soft gamma repeaters 
\cite{Kouveliotou,kulkarni,murakami} falls under such magnetars. Such magnetars show irregular spin-down rates, which are attributed to their precession and wobble due to strong magnetic fields. It was also recently suggested that the frame-dragging effect can get enhanced due to the presence of a strong magnetic field \cite{chandrachur2}. Therefore, it is essential to study both the geodetic and frame-dragging precession of a gyro for an NS and what effect do strong magnetic fields have on the gyro. 

Previously in the literature, there had been some studies regarding gyroscopic precession. The frame-dragging effect was initially investigated by lense and thrilling \cite{lt} and therefore, it is also known as lense thrilling (LT) effect. Hartle \cite{hartle} did the first calculation involving the frame-dragging effect in NS. Using perturbative approximation for slowly rotating stars, he calculated the structural changes in the star. Glendenning and Weber \cite{glen} showed that the LT effect could alter the Keplerian frequency and the Moment of Inertia of the star \cite{weber}. The problem of QPO was studied from this time by Morsink and Stella \cite{morsink} and they found that the QPO frequency is related with the oscillation frequency of the star along the $\theta$ and $\phi$ direction. Recent calculation of the LT effect of gyroscope near an NS was studied by Chandrachur et al. \cite{chandrachur1} and they also included a magnetic field in their calculation subsequently \cite{chandrachur2}.

In this paper, we consider both the frame-dragging and the geodetic effect on the gyroscope near a rotating NS. We also include a magnetic field in our calculation. In section II, we give the details of our formalism for the calculation of overall precession rate. Section III discusses the numerical code which we have used for calculating the values of metric coefficients and subsequently modeling the star. It also mentions the magnetic field configurations and the EoS that we have used for our calculation. Our results are given in section IV, and finally, in section V, we summarize our findings and conclude from them.

\section{Calculation of Overall Precession Frequency}
\begin{figure}
    \centering
    \includegraphics[width = 3.5in]{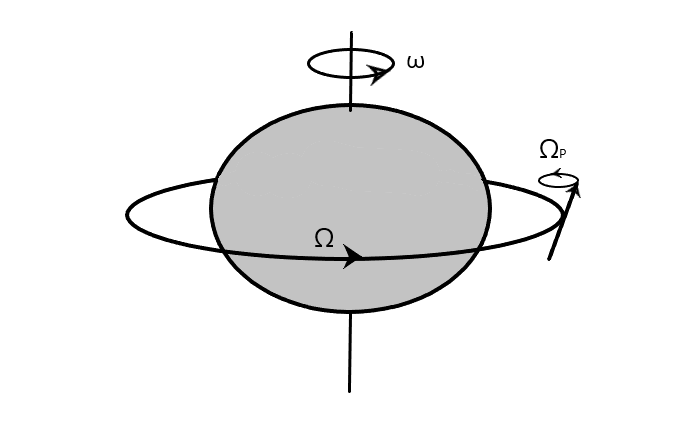}
    \caption{Sketch of a test gyro orbiting with orbital velocity $\Omega$ around a rotating NS whose angular velocity is $\omega$. The precession frequency of gyro is $\Omega_{P}$.}
    \label{precn}
\end{figure}

Our main objective is to obtain complete information about the geometry of spacetime by studying the variation of precession frequency of a test gyro orbiting in a circular orbit around an NS with constant nonzero angular velocity. 
The effect of curvature (geodetic) and the frame-dragging is revealed through the overall gyro precession frequency. The frame-dragging or the LT precession provides the information about rotation or magnetic field in the star.
Figure \ref{precn} shows the sketch of the gyro orbiting ($\Omega$) and precessing ($\Omega_P$) around an NS rotating with angular velocity $\omega$. In the case of NS, we can allow the orbits even in the interior of the star.  The time-like limit of the angular velocity of the gyro is finite and becomes more extensive as we move towards the center, which is shown in section IV. We can consider any spin vector as a test gyro, a spinning molecule or subatomic particle within the dense fluid of the NS. Thus studying the overall precession frequency of the gyro, we get the knowledge of the spacetime of a compact object.

\subsection*{Formalism}
The general spin precession frequency of a test gyro can be expressed in the form \cite{Straumann,Chakraborty:2016mhx}:   
\begin{eqnarray}
\tilde{\Omega}_P &=& \frac{1}{2K^2}*(\tilde{K} \wedge d\tilde{K})\ 
\label{b}
\end{eqnarray}
where $\Omega_P$ is the spin precession frequency in coordinate basis, $*$ represents the Hodge star operator or Hodge dual, $\wedge$ is the wedge product, $\tilde{K}$, $\tilde{\Omega}_P$ are the one-forms of killing vector $K$ and $\Omega_P$ respectively. In any stationary spacetime there exists a timelike killing vector. The metric of the spacetime discussed here is independent of $t$ and $\phi$. For a general stationary spacetime which also possesses a space-like Killing vector, we can write the general timelike Killing vector as a linear combination of the timelike killing vector $\partial_t$ and spacelike killing vector $\partial_\phi$ as follows:
\begin{eqnarray}
K=\partial_0+\Omega\ \partial_1.
\label{K}
\end{eqnarray}
where the coordinates (0,1) represent (t,$\phi$) and $\Omega$ represents the angular velocity for an observer moving along integral curves of $K$. The timelike killing vector approaches a unit timelike vector at spatial infinity. This is the reason we have kept the coefficient of $\partial_0$ to be 1. \\
The corresponding co-vector of $K$ is,
\begin{eqnarray}
\tilde{K}=g_{0\nu}dx^{\nu}+\Omega g_{\lambda 1}dx^{\lambda},
\end{eqnarray}
where $\lambda , \nu=0,1,2,3$ in 4-dimensional spacetime. 
Separating space and time components we can write $\tilde{K}$ as
\begin{eqnarray}
\tilde{K}=(g_{00}dx^0+g_{01}dx^1+g_{0i}dx^i) \nonumber
\\
+\Omega (g_{01}dx^0+g_{11}dx^1+g_{i1}dx^i)
\label{kt}
\end{eqnarray}
where $i=2,3$. Since we are mainly interested in a stationary and axisymmetric spacetime, the terms $g_{0i}$ and $g_{i1}$ are zero. Thus, we obtain
\begin{eqnarray}
\tilde{K}=(g_{00}dx^0+g_{01}dx^1)+\Omega (g_{01}dx^0+g_{11}dx^1)
\label{kto}
\end{eqnarray}
and 
\begin{eqnarray}
d\tilde{K}=(g_{00,k}dx^k \wedge dx^0+g_{01,k}dx^k \wedge dx^1) \nonumber
\\
+\Omega (g_{01,k}dx^k \wedge dx^0+g_{11,k}dx^k \wedge dx^1) .
\label{dkto}
\end{eqnarray}

The $\tilde{K} \wedge d\tilde{K}$ becomes

\begin{eqnarray}
 \tilde{K} \wedge d\tilde{K} = 
 \big [g_{00}g_{00,k} + \Omega  (g_{0k}g_{00,k} + g_{00}g_{01,k}) + \nonumber 
 \\ {\Omega^2} g_{01}g_{01,k} \big] 
 \cdot(dx^0 \wedge dx^k \wedge dx^0)  \nonumber 
 \\
 + \big [g_{01}g_{01,k} + \Omega  (g_{11}g_{01,k} + g_{01}g_{11,k}) 
+ {\Omega^2} g_{11}g_{11,k} \big]\nonumber 
\\
\cdot(dx^1 \wedge dx^k \wedge dx^1) \nonumber 
\\
 + \big [g_{00}g_{01,k} + \Omega  (g_{01}g_{01,k} + g_{00}g_{11,k}) 
+ {\Omega^2} g_{01}g_{11,k} \big]\nonumber 
\\
\cdot(dx^0 \wedge dx^k \wedge dx^1) \nonumber 
\\
 +\big [g_{01}g_{00,k} + \Omega  (g_{11}g_{00,k} + g_{01}g_{01,k})
+ {\Omega^2} g_{11}g_{01,k} \big]\nonumber 
\\
\cdot(dx^1 \wedge dx^k \wedge dx^0)
 \label{intmd}
\end{eqnarray}

We use properties of wedge product as follows, 
\begin{equation*}
(dx^0 \wedge dx^k \wedge dx^0) = (dx^1 \wedge dx^k \wedge dx^1) = 0 
\end{equation*}
and
\begin{eqnarray}
(dx^1 \wedge dx^k \wedge dx^0) = - (dx^0 \wedge dx^k \wedge dx^1)
\end{eqnarray}
Using the value of $K^2=g_{00}+2\Omega g_{01}+\Omega^2 g_{11}$, and $*(dx^0 \wedge dx^k \wedge dx^1)=\eta^{0k1l}g_{l\mu}dx^{\mu}
=-\frac{1}{\sqrt{-g}}\varepsilon_{k1l}g_{l\mu}dx^{\mu}$ in eqn. \ref{b}, 
the one-form of the precession frequency becomes
\begin{eqnarray}
\tilde{\Omega}_P&=&\frac{\varepsilon_{1kl}g_{l\mu}dx^{\mu}}{2\sqrt {-g}\left(1+2\Omega\frac{g_{01}}{g_{00}}
    +\Omega^2\frac{g_{11}}{g_{00}}\right)}. \nonumber \\
&&\left[\left(g_{01,k}
-\frac{g_{01}}{g_{00}} g_{00,k}\right)+\Omega\left(g_{11,k}
-\frac{g_{11}}{g_{00}} g_{00,k}\right)+ \right. \nonumber
\\
&&\left. \Omega^2 \left(\frac{g_{01}}{g_{00}}g_{11,k}
-\frac{g_{11}}{g_{00}} g_{01,k}\right) \right]
\label{main1}
\end{eqnarray}

The corresponding vector ($\Omega_P$) of the one-form $\tilde{\Omega}_P$ is:
\begin{eqnarray}
\Omega_P&=&\frac{\varepsilon_{1kl}}{2\sqrt {-g}\left(1+2\Omega\frac{g_{01}}{g_{00}}
    +\Omega^2\frac{g_{11}}{g_{00}}\right)}. \nonumber
\\
&&\left[\left(g_{01,k}
-\frac{g_{01}}{g_{00}} g_{00,k}\right)+\Omega\left(g_{11,k}
-\frac{g_{11}}{g_{00}} g_{00,k}\right)+ \right. \nonumber
\\
&&\left.\Omega^2 \left(\frac{g_{01}}{g_{00}}g_{11,k}
-\frac{g_{11}}{g_{00}} g_{01,k}\right) \right]\partial_l .
\label{elt}
\end{eqnarray} 

Using the properties $\varepsilon_{1k1} = 0$ and ${\varepsilon}_{123} = - \varepsilon_{132} = 1$ and setting coordinates ($2,3 = r,\theta$), expression \ref{elt} reduces to following form: 
\begin{eqnarray}  
\Omega_P&=& \Omega^{\theta} \partial_{\theta}+\Omega^r \partial_r 
\end{eqnarray}

In order to make numerical calculations for the overall precession frequency, we need to transform the above expression from the coordinate basis to the orthonormal `Copernican' basis. In this basis, with our choice of spherical  coordinates, $\Omega_{P}$ can be written as 
\begin{equation}
\vec{\Omega}_P=\sqrt{g_{rr}}\Omega^r \hat{r}+\sqrt{g_{\theta\theta}}\Omega^{\theta} \hat{\theta}
\label{k9}
\end{equation}

In a stationary and axisymmetric spacetime with coordinates $ t, r, \theta, \phi $, expression \ref{k9} becomes 
\begin{eqnarray}
\vec{\Omega}_P&=&\frac{1}{2\sqrt {-g}\left(1+2\Omega\frac{g_{t\phi}}{g_{tt}}+\Omega^2\frac{g_{\phi\phi}}{g_{tt}}\right)}. \nonumber
\\
&&\left[-\sqrt{g_{rr}}\left[\left(g_{t\phi,\theta}
-\frac{g_{t\phi}}{g_{tt}} g_{tt,\theta}\right)+\Omega\left(g_{\phi\phi,\theta}
-\frac{g_{\phi\phi}}{g_{tt}} g_{tt,\theta}\right) \right.\right. \nonumber
\\
&& \left.\left.+ \Omega^2 \left(\frac{g_{t\phi}}{g_{tt}}g_{\phi\phi,\theta}
-\frac{g_{\phi\phi}}{g_{tt}} g_{t\phi,\theta}\right) \right]\hat{r} \right. \nonumber
\\
&&\left. +\sqrt{g_{\theta\theta}}\left[\left(g_{t\phi,r}
-\frac{g_{t\phi}}{g_{tt}} g_{tt,r}\right) \right. \left.+\Omega\left(g_{\phi\phi,r}
-\frac{g_{\phi\phi}}{g_{tt}} g_{tt,r}\right) \right.\right. \nonumber
\\
&&\left.\left.+  \Omega^2 \left(\frac{g_{t\phi}}{g_{tt}}g_{\phi\phi,r}
-\frac{g_{\phi\phi}}{g_{tt}} g_{t\phi,r}\right) \right]\hat{\theta} \right] .
\label{main}
\end{eqnarray}
 Once we find the metric coefficients of the given astrophysical spacetime, we can use the above formula to find the overall precession frequency of the test gyro.

\subsection*{Frame Dragging Frequency}

For $\Omega=0$, the above equation reduces to LT precession frequency of the test gyro due to the rotation in any stationary and axisymmetric spacetime \cite{Straumann}.

\begin{eqnarray}
\vec{\Omega}_P|_{\Omega=0}=&&\frac{1}{2\sqrt {-g}}\left[-\sqrt{g_{rr}}\left(g_{t\phi,\theta}
-\frac{g_{t\phi}}{g_{tt}} g_{tt,\theta}\right)\hat{r}\right.
\nonumber
\\
&&\left.+\sqrt{g_{\theta\theta}}\left(g_{t\phi,r}-\frac{g_{t\phi}}{g_{tt}}
g_{tt,r}\right)\hat{\theta}\right],
\label{ltp0}
\end{eqnarray}

To apply this to an NS, we need the values of metric coefficients as a function of coordinate variables. First, we use a spherically symmetric metric to model a 
NS and find the nature of overall precession frequency around it. Then we use XNS code \cite{Bucciantini} first to model a rotating NS and then a magnetized NS to study the effects of rotation and magnetic field distinctly.

\section{XNS code, magnetic field and EoS}
In spherical coordinates the line element used in XNS code can be written as \cite{Bucciantini}:

\begin{equation}
ds^2=-\alpha^2 dt^2+\psi^4[dr^2+
r^2d\theta^2+r^2\sin^2\theta(d\phi+\beta^\phi dt)^2],
\label{mett}
\end{equation}
where $\alpha$ is called the \emph{lapse} function, $\beta^\phi$ is the
\emph{shift vector} and $\psi$ is the \emph{conformal factor}. The spatial part of the metric is conformally flat. 

The spherical coordinate system $x^\mu = (t, r, \theta, \phi) $ is chosen and the assumptions of \emph{stationarity} and \emph{axisymmetry} imply that all metric terms are only a function of $r$ and $\theta$.

The stress-energy tensor is given by:
\begin{equation}
T^{\mu\nu} = (e+p+b^2) u^\mu u^\nu - b^\mu b^\nu + (p + b^2/2) g^{\mu\nu},
\label{eq:grmhd}
\end{equation}
where $e$ is the total energy density, $p$ is the pressure, $u^\mu$ is the 4-velocity of the fluid, and $b^\mu := F^{*\mu\nu}u_\nu$ is the magnetic field as measured in the comoving frame, and $F^{\mu\nu}$ is the Faraday tensor (the asterisk indicates the dual).

The Einstein equations are solved by providing proper initial guess values (obtained from the TOV solution). XNS uses a self-consistent method and searches for axisymmetric equilibrium solutions of the compact stars in the presence of magnetic field and rotation. The metric terms and fluid parameters are derived at the same time. For the magnetic field, some free functions are defined, and the appropriate value of these functions are chosen to obtain the desired value of the magnetic field and configuration. The code solves for poloidal, toroidal, and twisted torus type of magnetic field. The details of the code can be found in the article by Bucciantini and Zanna \cite{Bucciantini}, and we do not discuss them explicitly here.

\begin{figure*}
\centering
\includegraphics[width = 3.5in,height=2.3in]{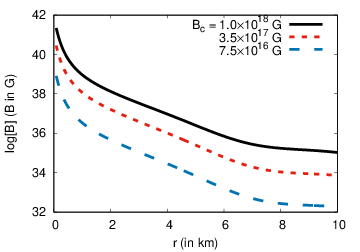}
\includegraphics[width = 3.5in,height=2.3in]{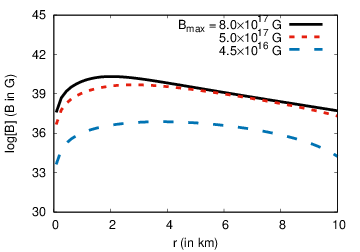}

\hspace{0.5cm} \scriptsize{(a)} \hspace{8.5 cm} \scriptsize{(b)} 
\caption{(Color online) The magnetic field configuration of the star is shown. Variation of the magnitude of the magnetic field with distance along the equatorial plane is shown for the poloidal (a) and toroidal (b) field. Curves are drawn for three different magnetic field strength, the black solid curve for the maximum strength and blue dashed curve for the minimal strength. The logarithmic value of the magnetic field is plotted along the y-axis as the field strength varies from $10^{14}$ G to $10^{18}$ G. The central energy density of star is taken to be $1.2\times10^{15}$ g $cm^{-3}$.}
\label{Bfield}
\end{figure*}

Figure \ref{Bfield} (a) shows the variation of the poloidal magnetic field inside a star. The maximum central value of the magnetic field is assumed to be $1.0 \times 10^{18}$ G, whereas at the surface its value decreases by four orders of magnitude. Figure 2(b) shows the variation of the toroidal magnetic field inside a star. The maximum central value is taken $8\times10^{17}$ G. Thus solving the XNS code we obtain the metric coefficients and also the magnetic field value in and around an NS.



\subsection*{EoS}

\begin{figure}
\centering
\includegraphics[width = 3.5in]{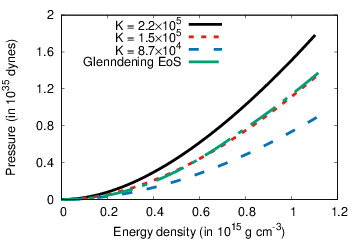}
\caption{Pressure, as a function of energy density, is plotted for four different EoS. The black solid curve describes the stiff polytropic EoS. The red dotted curve is for the polytrope with moderate stiffness. The blue dashed curve is for soft polytope, and the green long dashed curve is for the Glendenning EoS. The value of $\gamma$ for the polytropes is the same ($\gamma=2$) only the value of $K$ differs.}
\label{eos}
\end{figure}

The central object, the NS, is obtained by solving the XNS code. The code needs the EoS of matter to solve for a stable NS configuration. In this calculation, we have used Polytropic EoS described by the relation 
\begin{equation}
 P=K\rho^{\gamma}
\end{equation}
where P is the pressure, $\rho$ is the mass density, $\gamma$ is the polytropic index, and $K$ is the proportionality constant. In this calculation, we have kept the value of $\gamma=2$.
For comparison, we have also used a tabulated EoS with Glendenning parameter setting \cite{glen1}. The XNS code is challenging to solve with tabulated EoS, and therefore, we use mainly polytropes. This will not change our findings qualitatively. Fig. \ref{eos} shows the plot of four different EoS curves obtained from the Glendenning EoS and Polytropic EoS. We have plotted three curves with different values of $K$ to have quantitatively three different EoS, one soft, one standard and one stiff. The value of $\gamma$ for the curve remains the same, but we have changed the value of $K$ to obtain different slopes. $K=2.2\times10^5$ gives a stiff curve, whereas $K=8.7\times10^4$ gives a soft curve. The Glendenning EoS and $K=1.5\times10^5$ lie in between the two. Most of our calculation is done with the polytropic EoS with $K=1.5\times10^5$. The other polytropes with different $K$ values are to discuss the effect EoS have in the gyro precession.

\subsection*{Spherically Symmetric Metric}

The spherically symmetric metric is a good starting point to validate the results obtained using the XNS code for a non-rotating non-magnetic star.
Considering the general static and symmetric metric:
\begin{equation}
\label{metric}
d s^2 = \mbox{} - e^{\nu(r)} \, d t^2 +  e^{\lambda(r)} \, d r^2 + r^2 \,( d\theta^2 + sin^2\theta d\phi^2)
\end{equation}

Solving Einstein's equations for the values of $\nu$ and $\lambda$ using polytropic EoS, the interior solution is obtained by solving the following equations. 

\begin{equation} \label{dm} 
\frac{d m}{d r}  = 4 \pi r^2 \rho,
\end{equation}
\begin{equation}
\label{equati} 
\frac{d\nu}{d r}  = \frac{2 m(r) + 8\pi r^3p}{r[r - 2 m(r)]}, \\ e^{\lambda(r)} = \left[ 1 - \frac{2 m(r)}{r}\right] ^{-1}
\end{equation}
\begin{equation} \label{tov}
\frac{dp}{d r}  =  -(\rho + p) \frac{m(r) + 4\pi r^3 p}{r(r - 2 m(r))} .
\end{equation}

Eqns. \ref{dm} to \ref{tov} constitute the TOV equations or the equation of Hydrostatic equilibrium.

The metric coefficients outside the star are given by 

\begin{equation}
e^{\nu(r)} = \left[ 1 - \frac{2 M}{r}\right] ,
\end{equation}

\begin{equation}
e^{\lambda(r)} = \left[ 1 - \frac{2 M}{r}\right] ^{-1}. 
\end{equation}


Using the value of the metric coefficients and substituting it in eqn. \ref{main}, the vector $\vec{\Omega}_P$ in Copernican frame is given by
\begin{equation}
\vec{\Omega}_P = -\frac{\sqrt{e^{\nu }} \Omega  \cos \theta}{e^{\nu }-r^2 \Omega ^2 \sin ^2\theta } \hat{r} +
\frac{e^{\nu } \Omega  \sin \theta}{\sqrt{e^{\lambda +\nu }} \left(e^{\nu }-r^2 \Omega ^2 \sin ^2\theta\right)} \hat{\theta} .
\end{equation}
Thus the magnitude becomes
\begin{equation}
{\Omega}_P = \frac{\Omega  \sqrt{e^{\nu -\lambda } \left(e^{\lambda } \cos ^2\theta +\sin ^2\theta\right)}}{e^{\nu }-r^2 \Omega ^2 \sin ^2\theta }
\end{equation}

\begin{figure}
\centering
\includegraphics[width = 3.5in]{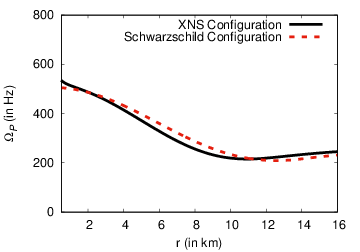}
\caption{Variation of $\Omega_{P}$ with distance from centre along the equatorial plane orbiting around spherically symmetric star with central energy density of $1.2\times10^{15} g cm^{-3}$ using Schwarzschild configuration and XNS configuration. The black solid curve is for static XNS configuration with $\beta_\phi=0$ in metric (see \ref{mett}) and the red dotted curve is for Schwarzschild configuration with ideal fluid interior.}
\label{sch}
\end{figure}

The variation of $\Omega_P$ with distance along the equatorial plane is shown in Figure \ref{sch}. Here a polytropic EoS with $K = 1.5\times10^5$ is used to build the NS. The precession frequency is maximum at the center, decreases to a local minimum near the surface of the star and later increases and saturates to the value of the orbital angular velocity of the gyro at a significant distance. The Schwarzschild and the XNS code results match reasonably. 

\subsection*{Axisymmetric Metric}
For a rotating or a magnetized NS, we calculate $\Omega_P$ by inserting the values of the metric coefficient in eqn \ref{main}. The $\Omega_P$ reduces to the simple form
\begin{equation}
\vec{\Omega}_P=\omega_r \hat{r}+\omega_{\theta} \hat{\theta}
\label{k91}
\end{equation}
where
\begin{equation}
\omega_r = -\frac{{\psi} A \left(\psi  \left(\frac{\partial\beta}{\partial\theta} B-C \frac{\partial\alpha}{\partial\theta}\right)+\alpha ^2 (2 \beta +\Omega ) \frac{\partial\psi}{\partial\theta}\right)}{r^2 \psi ^5 \left(2 \beta ^2 r^2 \psi^2-\alpha ^2\right) D}
\label{wr}
\end{equation}
\begin{equation*}
\omega_{\theta} = \frac{r \psi E \left(\psi  \left(r \frac{\partial\beta}{\partial r} F + G\right)+\alpha ^2 r (2 \beta +\Omega ) \frac{\partial\psi}{\partial r}\right)}{r^3 \psi ^5
    \left(2 \beta ^2 r^2 \psi ^2-\alpha ^2\right) H}
\end{equation*}
\begin{equation*}
A =  \sqrt{\alpha ^2 r^4 \psi ^6-2 \beta ^2 r^6 \psi ^8}
\end{equation*}
\begin{equation*}
B = \left(\alpha ^2+r^2 \psi ^2 \left(2 \beta  (\beta +\Omega )+\Omega
^2\right)\right)
\end{equation*}
\begin{equation*}
C = \alpha  (2 \beta +\Omega )
\end{equation*} 
\begin{equation*}
D = \left(r^2 \psi ^2 \left(2 \beta  (\beta +2 \Omega )+\Omega ^2\right)-\alpha ^2\right)
\end{equation*}
\begin{equation*}
E = \sqrt{\alpha ^2 r^4 \psi ^6-2 \beta ^2 r^6 \psi ^8}
\end{equation*}
\begin{equation*}
F = \left(\alpha ^2+r^2 \psi ^2 \left(2 \beta  (\beta +\Omega) +\Omega^2\right)\right)
\end{equation*}
\begin{equation*}
G = \alpha  (2 \beta +\Omega ) \left(\alpha -r \frac{\partial\alpha}{\partial r}\right)
\end{equation*}
\begin{equation*}
H = \left(r^2 \psi ^2 \left(2 \beta  (\beta +2 \Omega )+\Omega ^2\right)-\alpha ^2\right).
\end{equation*}

The graphical description of the variation of $\lvert{\vec{\Omega}_P}\rvert$ is shown in the subsequent section.

\section{Results}

We calculate the precession frequency of the test gyro in and around a rotating or a magnetized NS. We analyze the gyro frequency, and from the nature of the gyro precession frequency (GPF), we calculate the properties of the central object. First, we want to analyze if the GPF can distinguish between an NS or a BH. Further, we analyze the nature of GPF for an NS having different mass, frequency, EoS, and magnetic field. We have plotted results with the star solution obtained by solving the XNS code. The results and the figures are plotted mostly along the equatorial plane unless and otherwise stated. The results are shown for polytrope with $K=1.5\times 10^5$ unless stated otherwise.

\subsection*{Range of $\Omega$}

Determining the values of the metric coefficients, we proceed further to study the properties of spacetime and behavior of test gyro. For four-velocity $u$ to be time-like, the killing vector must be negative and is given by
\begin{eqnarray}
K^2 = g_{\phi\phi}\Omega^2+2g_{t\phi}\Omega+g_{tt} & < & 0,
\end{eqnarray} 

Solving the above equation one obtains the allowed range of $\Omega$ at any fixed $(r,\theta)$,
\begin{eqnarray}
\Omega_-(r,\theta) < \Omega(r,\theta) < \Omega_+(r,\theta).
\label{oeg}
\end{eqnarray}

\begin{figure}
\centering
\includegraphics[width = 3.5in]{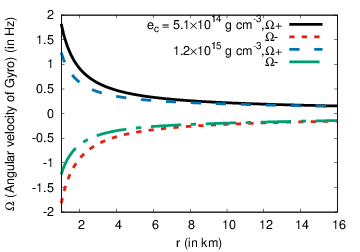}
\caption{Plots of the limiting values of the angular velocity of the gyro with distance from the center of the NS. The black solid curve for prograde motion and red short dashed curve for retrograde motion are for central energy density $5.1\times10^{14} g cm^{-3}$ and the blue long dashed curve for prograde motion and green solid-dashed curve for prograde motion for central energy density of $1.2\times10^{15} g cm^{-3}$. }
\label{rg1}
\end{figure}

Figure \ref{rg1} gives a range of $\Omega$, which satisfies the time-like limit. The range of $\Omega$ gets narrower as we move away from the center of the star. Also for massive stars, the range is more restricted than for a less massive star. Therefore, the value of $\Omega$, which we choose for our calculation lies within this range.

\subsection*{BH or NS}

In this section, we study the gyro precession frequency and obtain the difference the gyro frequency would suffer depending on the central object. 

\begin{figure*}
\centering
\includegraphics[width = 3.0in,height=2.3in]{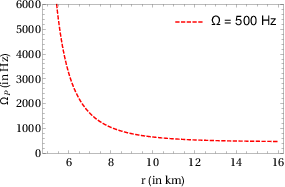}
\includegraphics[width = 3.5in,height=2.3in]{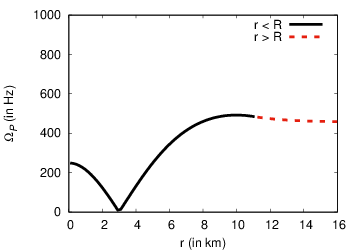}

\hspace{0.5cm} \scriptsize{(a)} \hspace{8.5 cm} \scriptsize{(b)} 
\caption{(Color online) (a)Variation of the GPF along the equatorial plane around a $1.5 M_0$ Kerr BH ($a = 0.1 M_0$) with $\Omega = 500 Hz$. (b) GPF along the equatorial plane around a $1.5 M_0$ NS of radius  $\sim 11$ km with $\Omega = 500 Hz$. The black curve and the red dotted curve are for interior and exterior regions of the star respectively.}
\label{ac1}
\end{figure*}

 The Kerr metric is used for describing a spinning BH and in Boyer-Lindquist coordinates it is written as,
\begin{eqnarray}
ds^2&=&-\left(1-\frac{2Mr}{\rho^2}\right)dt^2 - \frac{4Mar\sin^2\theta}{\rho^2}d\phi dt + \frac{\rho^2}{\Delta}dr^2 . \nonumber
\\
&& + \rho^2 d\theta^2 + \left(r^2+a^2+\frac{2Mra^2 \sin^2\theta}{\rho^2}\right) \sin^2\theta d\phi^2 
\label{k1}
\end{eqnarray}
where, $a$ is the Kerr parameter, defined as $a=\frac{J}{M}$, 
the angular momentum per unit mass and

\begin{equation*}
\rho^2=r^2+a^2 \cos^2\theta,
\label{k2}
\end{equation*}
\begin{equation*}
\Delta=r^2-2Mr+a^2.
\end{equation*}

We find overall precession frequency by putting the metric coefficients in equation \ref{main} and it comes out to be
\begin{equation}
|\vec{\Omega}_P| = {\frac{\left(X+2 a M (a \Omega  (a \Omega -1)+1)+r^3 \Omega \right)}{r^2 \left(r \left(\Omega ^2 \left(a^2+r^2\right)-1\right)+Y\right)}}
\end{equation}
where, 
\begin{equation*}
X = 3 M r^2 \Omega  (2 a \Omega -1)
\end{equation*}
and
\begin{equation*}
Y = 2 M (a \Omega  (a \Omega -4)+1)
\end{equation*}

In fig \ref{ac1} (a) we draw the curve for the central object being a BH and in Fig \ref{ac1} (b) the curve is drawn with the central object being an NS. The curves are drawn for azimuthal angle $\theta=90^{\circ}$, i.e., along the equatorial direction.
The curve shows the difference in the behavior of $\Omega_P$ with distance from the center along the equatorial plane for Kerr BH and NS. For a BH with smaller values of $\Omega$ (500 Hz), the behavior of $\Omega_P$ is monotonic and diverges as we approach the ergosphere. Whereas for an NS, the GPF is almost constant as we approach the NS. However, inside the NS GPF dips and goes to zero around $3$ km and increases again as we move towards the center. Thus seeing the behavior of spinning test gyro approaching a compact object, the central object can be identified whether it is a BH or an NS.

\subsection*{Dependence of precession frequency on the orbital angular velocity of gyro}

\begin{figure}
\centering
\includegraphics[width = 3.5in]{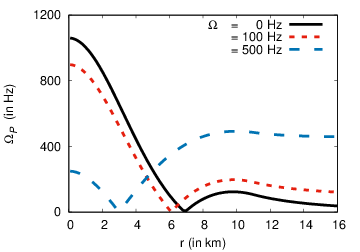}
\caption{LT precession frequency and overall precession rate along the equatorial plane around a $1.5 M_0$ NS ($\omega = 700 Hz$) of radius  $11$ km with varying the orbital angular velocity of the gyro. The black solid curve denotes the LT precession rate, the red dotted curve and the blue dashed curve denote overall precession rate for $\Omega = 100 Hz$ and $\Omega = 500 Hz$ respectively. There is a null region for all curves.}
\label{1o90}
\end{figure}

\begin{figure}
\centering
\includegraphics[width = 3.5in]{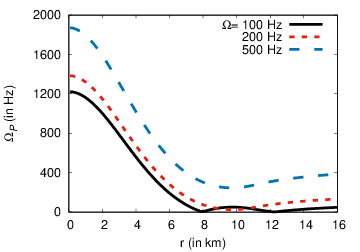}
\caption{Overall precession frequency along the equator for a $1.5 M_0$ NS ($\omega = 700 Hz$) with a sense of orbital angular velocity of the gyro opposite to that of star's rotation. This is a unique case for NS as there are two null regions for the precession curve. The black solid curve is for $\Omega=100 Hz$, the red dotted curve is for $\Omega=200 Hz$, and the blue dashed curve is for $\Omega=500 Hz$.}
\label{nao90}
\end{figure}

If the spin of the gyro to be zero, the overall frequency becomes the same as the LT frequency. The LT precession frequency at a particular distance from center depends on the change in frame-dragging term of the metric.
The drag is maximum at the center and decreases outwards. Near the center, the change in frame-dragging is negative. This is clear when we look into the terms in equation \ref{main}. The two dominating terms competing with each other are $\frac{dg_{t\phi}}{dr} $ ($negative$) and  $ \Omega \frac{dg_{\phi\phi}}{dr} $ ($positive$). They compete with each other to determine the overall precession rate of the gyro. Initially, the first term dominates, and we have an overall negative value. However, with an increase in radial distance, the positive term starts to increase (fig \ref{1o90}).  
This makes the overall negative value to decrease. At about 6 km the two-term becomes equal, and so the overall frequency becomes zero. Beyond that point, the positive term increases till 9 km, and we obtain maxima around 9 km. Further radially outwards both the term decreases and asymptotically the GPF becomes equal to the actual frequency of the gyroscope. 
As we are plotting only the magnitude, the curve always lies in the positive quadrant.

The overall frequency is obtained when the gyro has non zero spins. In this case, the magnitude of the resultant is shown by the red dotted curve, as shown in Fig \ref{1o90}. The angular velocity of the gyroscope is positive with respect to the angular velocity of the star (prograde motion). Therefore, the initial values of the overall precession frequency of the gyroscope decrease with increasing $\Omega$ as both effects have opposite sign. As $\Omega$ increases the initial overall term becomes less negative as the positive term increases. Thus in figure \ref{1o90}, the gyro with the highest angular velocity ($500 Hz$) has the least overall precession rate near the center, and also the minima occur at the smaller radial point from the center.

If the sense of revolution of the gyro is opposite to that of the angular velocity of the star (retrograde), the precession rate becomes more negative.  This is because the frame-dragging rate and other effects (which includes the geodetic effect) have the same sign. Therefore the initial values of overall precession frequency of the gyroscope increase with increasing $\Omega$.   In Figure \ref{nao90}, we plot the for the retrograde motion of the gyro. In case of an extreme orbital frequency like for example $\sim 500 Hz$ shown in figure \ref{nao90}, the nature of $\Omega_P$ is coincidentally similar to that of a static star but with opposite signs. 

$\Omega_P$ for the prograde motion of the gyro has single minima and maxima inside the star because as we move outward, the sign of dominating terms change only once. But in case of retrograde angular velocity, the values become more negative and change sign twice. Ultimately it settles to a negative $\Omega$ as we move outside the star (however, the plots shown is positive as we have plotted the magnitude of $\Omega_P$). But when the value of retrograde $\Omega$ is sufficiently large, the values change sign only once. The plot has a single extremum, which is a minimum in the interior of the star and has no local extremum, as shown in Figure \ref{nao90}. 

\subsection*{Dependence on azimuthal position}

\begin{figure}
\centering
\includegraphics[width = 3.5in,height=2.3in]{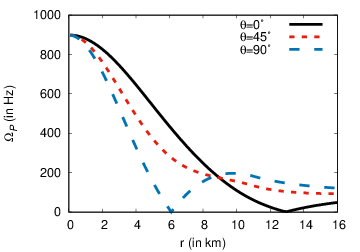}

\caption{(Color online)(a) $\Omega_P$ of gyro along different azimuthal positions around a $1.5 M_\circ$ NS ($\omega=700 Hz$) with positive orbital angular velocity $\Omega=100 Hz$. The black solid curve, red dotted curve and blue dashed curve are for $\theta = 0^{\circ},45^{\circ},90^{\circ}$ respectively.}
\label{agn}
\end{figure}

In Figure \ref{agn} the azimuthal dependence of the GPF is plotted. Along the equator, there is one null point (consistent with Fig \ref{1o90}). As we move towards smaller azimuthal angles first, the null point disappears and then the minima is also lost at $\theta=45^{\circ}$. Moving further to much lower angles the minima again reappears and finally a null point is generated for the pole.
The position of the polar null point of the gyro precession occurs at the larger radial distance than that of an equatorial null point.

\subsection*{Dependence on the angular velocity of the NS}

\begin{figure}
\centering
\includegraphics[width = 3.5in,height=2.3in]{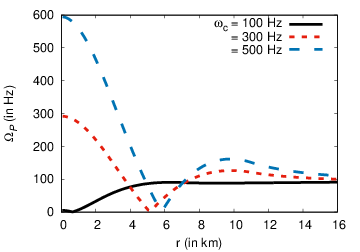}

\caption{(Color online) (a) Effect of the angular velocity of the star on $\Omega_P$ around a $1.5 M_\circ$ NS with central energy density $1.2\times10^{15} g cm^{-3}$. The solid black line, red dotted line, and  blue dashed line are for $\omega_c = 500 Hz, 300 Hz, 100 Hz$ respectively.}
\label{all}
\end{figure}
Keeping the angular velocity of the gyro to be $\Omega=100$ Hz, we find that the gyro precession goes to zero at the center of the star having an angular velocity of 100 Hz ($\omega$). This is a robust result, and whenever both the angular velocity becomes equal, the gyro precession at the center of the star becomes zero. As we move outwards, it starts to grow and asymptotically attains a value of 100 Hz. Increasing the angular velocity of the star increases the drag in and around the central object. This frame-dragging rate increases the overall precession rate proportionally, as shown in Figure \ref{all}.  

\subsection*{Dependence on Mass and EoS}

\begin{figure*}
\centering
\includegraphics[width = 3.5in,height=2.3in]{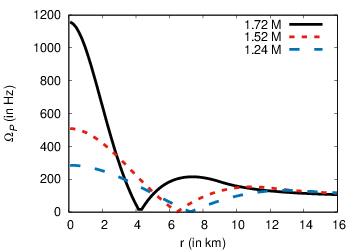}
\includegraphics[width = 3.5in,height=2.3in]{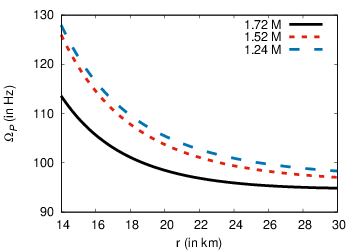}

\hspace{0.5cm} \scriptsize{(a)} \hspace{8.5 cm} \scriptsize{(b)} 
\caption{(Color online) (a) $\Omega_P$ of the gyro around NS of different mass is shown in the figure. The black solid line shows variation of $\Omega_P$ for a $1.72 M_\circ$ star, the red dotted and blue dashed solid line are for $1.52 M_\circ$ and $1.24 M_\circ$ respectively. (b) $\Omega_P$ of the gyro in the exterior region of the star. The nomenclature of curves are same as that of (a).}
\label{mn}
\end{figure*}

In Fig \ref{mn} (a), we have plotted the gyro overall precession frequency as a function radial distance from the center of the star. The three different curves are for three different mass configuration having $\omega=500 Hz$ and $\Omega=100 Hz$. We have plotted curves along the equatorial direction. The NS having higher mass, the precession frequency of the gyro revolving around it will be higher. Keeping all rotation effects constant, increasing the central density increases the mass, which increases the change in temporal and radial coefficients. This increases the value of initial precession frequency and the maxima. In fig \ref{mn} (b) we have plotted GPF beyond the surface of the NS. The nature of the precession frequency remains the same; however, there is some difference in the overall value. Therefore, if the GPF near an NS can be measured accurately, then one can get some hint of the mass of an NS around which it is precessing. 

\begin{figure}
\centering
\includegraphics[width = 3.5in,height=2.3in]{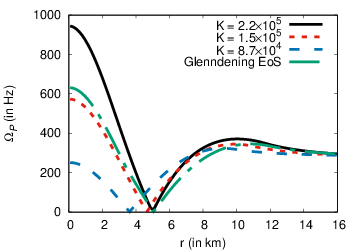}

\caption{(Color online) $\Omega_P$ of the gyro around NSs of four different EoS is shown. The black solid line is for a stiff polytropic star. The red dotted curve is for a standard polytropic star. The blue-dashed line is for a softer polytropic EoS and the green dot dashed curve is for Glendenning EoS.}
\label{en}
\end{figure}

The maximum mass and the compactness of an NS depends on the EoS describing the star. Stiff EoS gives rise to more compact and massive stars, whereas soft EoS offers less massive stars. If the EoS of the star is stiff, it rapidly changes the values of metric coefficients, which leads to the shift in the gyroscopic precession frequencies. Figure \ref{en} shows that the stiffer the EoS more the value of GPF at the center of the star. However, the behavior of the curves beyond the star radius is more or less identical. Therefore, it is difficult to differentiate stars of the same mass having different EoS using a gyro.

\subsection*{Dependence on nature and strength of magnetic field}

\begin{figure*}
\centering
\includegraphics[width = 3.5in,height=2.3in]{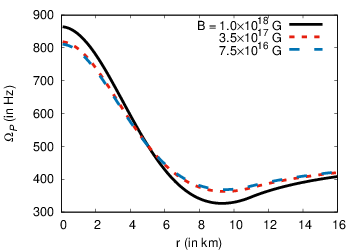}
\includegraphics[width = 3.5in,height=2.3in]{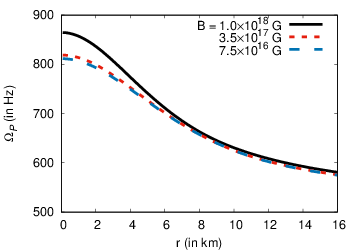}

\hspace{0.5cm} \scriptsize{(a)} \hspace{8.5 cm} \scriptsize{(b)} 
\caption{(Color online) We have plotted the effect of a poloidal magnetic field on $\Omega_P$ along the equatorial plane (a) and polar plane (b). The black solid curve, the red dotted curve and the blue dashed curve are for central magnetic field of $1.0\times10^{18} G, 3.5\times10^{17} G, 7.5\times10^{16} G$ respectively. $\Omega$ is taken to be $500 Hz$. The central energy density of the NS is taken to be $1.2\times10^{15} g cm^{-3}$.}
\label{bmag}
\end{figure*}

Here we show the effect of the magnetic field on $\Omega_P $ of the test gyro. To discuss the impact of the magnetic field quantitatively, we have made the frame-dragging effect to be zero (non-rotating NS). Here the GPF has no local maxima inside the star unlike for most of the previous cases with frame-dragging effect. The initial value of the precession frequency is higher for the higher values of the central magnetic field for the poloidal field, as can be seen from Fig \ref{bmag} (a). Outside the star, the difference in the GPF for different field strength is minimal, and it is difficult to test the magnetic field strength of a star using a gyro.

In fig \ref{bmag} (b) we have plotted the GPF along the polar plane.
Along the equatorial direction, the curves are stiffer than along the polar plane as the variation of the poloidal field along the equator is much stiffer. Along the equatorial plane, the curves have a local minimum, whereas for the polar plane the curves decrease monotonically. Along the equatorial direction, the deformation is highest as the star becomes oblate. The change in metric coefficients is largest and thus explaining the plots in Figure \ref{bmag} (a). The effect of the magnetic field becomes significant only after the central value of the field is a few times $10^{17}$ G.

\begin{figure*}
\centering
\includegraphics[width = 3.5in,height=2.3in]{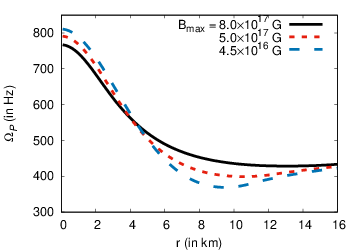}
\includegraphics[width = 3.5in,height=2.3in]{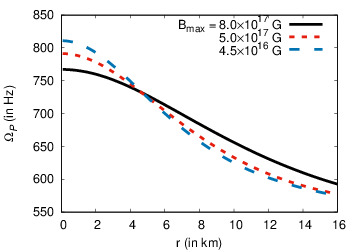}

\hspace{0.5cm} \scriptsize{(a)} \hspace{8.5 cm} \scriptsize{(b)} 
\caption{(Color online) We have plotted the effect of a toroidal magnetic field on $\Omega_P$ along the equatorial plane (a) and along the polar plane (b). The black solid curve, the red dotted curve and the blue dashed curve are for central magnetic field of $8.0\times10^{17} G, 5.0\times10^{17} G, 4.5\times10^{16} G$ respectively. $\Omega$ is taken to be $500 Hz$. The central energy density of the NS is taken to be $1.2\times10^{15} g cm^{-3}$.}
\label{bmagt}
\end{figure*}

However, in case of the toroidal field, the star becomes prolate 
and the curves become drastically different. The initial value of the precession frequency is higher for the lowest values of the central magnetic field, as can be seen from Fig \ref{bmagt}. The curve with the largest magnetic field strength is flatter than the curve with less field strength. The GPF outside the star for three different magnetic field strength is also not much different. Along the polar direction, the curves become flatter with increase in magnetic field without any local minima. 

For the poloidal case, the magnetic field makes the gyro precession curve stiffer, and the maximum of the gyro frequency increases at the center. As the curves become stiffer, the local minima become more prominent. However, for the toroidal field, the magnetic field makes the gyro frequency curve flatter, and the maxima of the gyro frequency at the center goes down. As the field strength grows, the curves become flatter, and the local minima disappear. This is more prominent inside the star, but also have some effect near the surface. The change in the magnetic field distribution changes the nature of the temporal component of the metric ($g_{tt} = -\alpha^2$) and also the radial and azimuthal distribution of mass in the star. The poloidal and toroidal magnetic field distribution are shown in figures 2 (a and b). Looking at the nature of the GPF, we can at least get some hint of the magnetic field distribution inside the star.

Looking at the overall GPF as we approach an NS (or a BH) we can differentiate between central objects (whether it an NS or a BH). For an NS, we can scrutinize its mass and the nature of the magnetic field. Most of the characteristics are prominent inside the star, but some of them have considerable effect near the surface.

\section{Summary and Conclusion}

In this work, we had studied the effect of an NS on a test gyroscope when placed near it. The study is essential to track the relativistic effect that affects a particle in the field of NS gravity. If the star is rotating the relativistic effect is that of geodetic and frame-dragging. For a magnetized star both the geodetic precession (affecting the overall curvature) and the frame-dragging precession are important.

The GPF includes both the geodetic and the frame-dragging effect. The geodetic effect gives the outcome of the curvature, and the LT precession provides the information about rotation or magnetic field in the star. The local minima of the GPF depend on the angular distribution and the orbital velocity of the gyroscope. The star's compactness (mass and radius) also plays a vital role in determining the overall precession of the star. 

The GPF behaves differently when it approaches a BH or an NS. The GPF diverges as it approaches a BH, whereas, for an NS, it is always finite. Therefore, a particle would behave differently when it approaches a BH or an NS. For a BH the minima are not always present, whereas for an NS it is always there along the equatorial plane.
Thus, the GR effect on a particle differs depending on whether it is in the proximity of a BH or inside an NS.

The angular frequency of the gyro is fundamental in determining the gyro precession frequency. If the gyroscope has a prograde motion with the star, the central value of precession frequency is maximum for a static gyro and decreases with increase in gyro angular velocity. However, if the gyro is in retrograde motion, the gyro frequency increases with an increase in gyro angular velocity. Also, the number of null points differ for prograde and retrograde motion. The location of the null point of the gyro precession frequency also depends on the azimuthal position of the NS. This is important in determining the movement of a real particle encircling an NS.

The angular velocity of the NS is also essential towards determining the GPF. If the gyro and the NS have the same angular velocity, then the value of GPF at the center of the star is zero and increases monotonically with radial distance from the center. However, if the gyro and NS angular velocity are different, then the GPF rises with an increase in NS angular velocity. Also, the null point shifts towards the surface with an increase in angular velocity of the NS. The GPF also changes with the change in the mass of the star. Therefore, observing the GPF, we can get some idea about the angular velocity and the mass of an NS.

In the present calculation, we have used two different field configuration, poloidal field, and toroidal fields in the star. The precession rate and the occurrence of the local minima depend strongly on the distribution of the magnetic field. For poloidal fields, an increase in the field strength makes the GP curve stiffer, and the local minima are much more prominent. Whereas, for the toroidal fields, the growth in fields strength makes the GP curve flat and the local minima disappear. Therefore, we can safely conclude that looking at the GP; we can distinguish between the magnetic field profile distribution present in the star. We can also deduce what GR forces a real particle would feel inside the star and what is the general relativistic effect that is important there.

We should mention here that we have done our calculation with a gyroscope having zero mass. However, for real-life scenario particle has some definite mass. The mass of the orbiting particle can be thought to of negligible mass if we consider a massive NS. The assumption is not inappropriate, but we can improve our results if we include a mass term in our calculation. 
From the above discussion, we conclude that the gyro can differentiate some properties of an NS as it approaches it. However, our test gyros can only be particles encircling an NS, like particles in an accretion disk. If the precession frequency of such particles can be inferred from astrophysical observation, we can deduce some properties on NS around which it is encircling.
Once the general relativistic effects near an NS are identified, we can then move to do more relevant problems like accretion in NS, the motion
of binary NS and the tilt in the accretion disk of NS. All this effect can have critical observational signatures, and recent astrophysical observation is bound to provide such trademarks. 

\acknowledgments
The authors are grateful to IISER Bhopal for providing all the research and infrastructure facilities. RM would also like to thank the SERB, Govt. of India for monetary support in the form of Ramanujan Fellowship (SB/S2/RJN-061/2015) and Early Career Research Award (ECR/2016/000161).

\end{document}